\documentclass[pre,aps]{revtex4}
\usepackage{graphicx}

\begin{document}

\title{Thermodynamics of classical frustrated spin chain at the
ferromagnet-helimagnet transition point}
\author{D.~V.~Dmitriev}
\email{dmitriev@deom.chph.ras.ru}
\author{V.~Ya.~Krivnov}
\affiliation{Joint Institute of Chemical Physics, RAS, Kosygin str.
4, 119334, Moscow, Russia.}
\date{}

\begin{abstract}
Low-temperature thermodynamics of the classical frustrated ferromagnetic
spin chain is studied. Using transfer-matrix method we found the behavior of
the correlation function and zero-field susceptibility at the
ferromagnetic-helical transition point. It is shown that the critical
exponent for the susceptibility is changed from 2 to 4/3 at the transition
point.
\end{abstract}

\maketitle

Lately, there has been considerable interest in low-dimensional spin models
that exhibit frustration. One of them is the spin chain with the
ferromagnetic interaction $J_{1}$ of nearest neighbor (NN) spins and the
antiferromagnetic next-nearest-neighbor (NNN) interaction $J_{2}$, so called
the 1D F-AF model. Its Hamiltonian has a form
\begin{equation}
H=J_{1}\sum \mathbf{S}_{n}\mathbf{\cdot S}_{n+1}+J_{2}\sum \mathbf{S}_{n}%
\mathbf{\cdot S}_{n+2}  \label{H}
\end{equation}%
where $J_{1}<0$ and $J_{2}>0$.

This model is characterized by a frustration parameter $\alpha
=J_{2}/|J_{1}| $. The ground state properties of the quantum $s=1/2$
F-AF chain have been intensively studied last years
\cite{Chubukov,DK06,HM,Hikihara}. It is known that the ground state
of the model is ferromagnetic for $\alpha <1/4$. At $\alpha =1/4$
the ground state phase transition to the incommensurate singlet
phase with helical spin correlations takes place. Remarkably, this
transition point does not depend on a spin value, including the
classical limit $s=\infty $.

Interesting question is the influence of the frustration on the
low-temperature thermodynamics of the model especially near the transition
point $\alpha =1/4$. We study this problem for the classical version of
model (\ref{H}). At zero temperature the classical model has long
range-order (LRO) for all values of $\alpha $: the ferromagnetic LRO at $%
\alpha \leq 1/4$ and the helical one at $\alpha >1/4$. At finite temperature
the LRO is destroyed by thermal fluctuations and thermodynamic quantities
have a singular behavior at $T\rightarrow 0$. In particular, the zero-field
susceptibility $\chi $ diverges. For the 1D Heisenberg ferromagnet ($\alpha
=0$) $\chi =2\left\vert J_{1}\right\vert /3T^{2}$ \cite{Fisher}. At $%
0<\alpha <1/4$ the susceptibility is $\chi =2(1-4\alpha )\left\vert
J_{1}\right\vert /3T^{2}$. This behavior of $\chi $ is similar to that for
the quantum $s=1/2$ F-AF model \cite{Hartel}. The prefactor in $\chi $
vanishes at the transition point indicating the change of the critical
exponent. We focus our attention on the behavior of $\chi $ at the
transition point.

The partition function $Z$ of model (\ref{H}) at $\alpha =1/4$ is
\begin{equation}
Z=\prod_{n=1}^{N}\int d\Omega _{n}\exp \left\{ \frac{1}{T}\sum (\vec{S}%
_{n}\cdot \vec{S}_{n+1}-\frac{1}{4}\vec{S}_{n}\cdot \vec{S}_{n+2})\right\}
\label{Z1}
\end{equation}%
where $\vec{S}_{n}$ is unit vector, $d\Omega _{n}$ is the volume element of
the solid angle for $n$-th site, we put $\left\vert J_{1}\right\vert =1$ and
the periodic boundary conditions are proposed.

Our further calculations are based on the transfer matrix method and we use
a version of this method adapted to the model with NNN interactions by
Harada and Mikeska in \cite{Harada-Mikeska}.

Following Ref.\cite{Harada-Mikeska} we represent $Z$ in a form
\begin{equation}
Z=\prod_{n=1}^{N}\int d\Omega _{n}K(\theta _{n-1},\theta _{n};\varphi _{n})
\label{Z2}
\end{equation}%
where
\begin{equation}
K(\theta _{n-1},\theta _{n};\varphi _{n})=\exp \left( \frac{\cos \theta
_{n-1}+\cos \theta _{n}}{2T}-\frac{(\cos \theta _{n-1}\cos \theta _{n}+\sin
\theta _{n-1}\sin \theta _{n}\cos \varphi _{n})}{4T}\right)   \label{K}
\end{equation}%
where $\theta _{n}$ is the angle between $\vec{S}_{n}$ and $\vec{S}_{n+1}$
and $\varphi _{n}$ is the angle between components of $\vec{S}_{n-1}$ and $%
\vec{S}_{n+1}$ projected onto $(X_{n},Y_{n})$ plane of the $n$-th local
coordinate system with the $Z_{n}$ axis parallel to $\vec{S}_{n}$.

Integrating Eq.(\ref{K}) over $\varphi _{n}$ we obtain $Z$ in a form
\begin{equation}
Z=\prod_{n}\int_{0}^{\pi }d\theta _{n}\sin \theta _{n}A(\theta _{n-1},\theta
_{n})  \label{Z3}
\end{equation}%
where
\begin{equation}
A(\theta _{n-1},\theta _{n})=\frac{1}{2}I_{0}(-z)\exp \left( \frac{\cos
\theta _{n-1}+\cos \theta _{n}}{2T}-\frac{\cos \theta _{n-1}\cos \theta _{n}%
}{4T}-\frac{3}{4T}\right)   \label{A1}
\end{equation}%
and $I_{0}(-z)$ is the modified Bessel function of%
\begin{equation}
z=\frac{\sin \theta _{n-1}\sin \theta _{n}}{4T}  \label{z}
\end{equation}

Let us consider an integral equation
\begin{equation}
\int_{0}^{\pi }A(\theta _{1},\theta _{2})\psi _{\alpha }(\theta _{2})\sin
\left( \theta _{2}\right) d\theta _{2}=\lambda _{\alpha }\psi _{\alpha
}(\theta _{1})  \label{I1}
\end{equation}%
where $\psi _{\alpha }(\theta )$ satisfy normalization condition
\begin{equation}
\int_{0}^{\pi }\psi _{\alpha }(\theta )\psi _{\beta }(\theta )\sin \theta
\mathrm{d}\theta =\delta _{\alpha ,\beta }  \label{norma}
\end{equation}

Eigenfunctions $\psi _{\alpha }$ and eigenvalues $\lambda _{\alpha }$ can be
chosen as real, since the kernel $A(\theta _{1},\theta _{2})$ is real and
symmetric. Then,
\begin{equation}
A(\theta _{1},\theta _{2})=\sum_{\alpha }\lambda _{\alpha }\psi _{\alpha
}(\theta _{1})\psi _{\alpha }(\theta _{2})  \label{A2}
\end{equation}

Substituting Eq.(\ref{A2}) into Eq/(\ref{Z3}) we obtain in the thermodynamic
limit
\begin{equation}
Z=\lambda _{0}^{N}  \label{Z4}
\end{equation}%
where $\lambda _{0}$ is the largest eigenvalue of Eq.(\ref{I1}).

In the low-temperature limit the angles $\theta _{n}$ are small and we can
use the asymptotic expansion of the modified Bessel function
\begin{equation}
I_{0}(-z)=\frac{e^{z}}{\sqrt{2\pi z}}\left( 1+\frac{1}{8z}+O(z^{-2})\right)
\label{I0z}
\end{equation}

Then, we expand the expression in the exponent of the transfer matrix to the
fourth order in $\theta _{i}$ to obtain
\begin{equation}
A(\theta _{1},\theta _{2})=\sqrt{\frac{T}{2\pi \theta _{1}\theta _{2}}}%
\left( 1+\frac{T}{2\theta _{1}\theta _{2}}\right) \exp \left( -\frac{(\theta
_{1}-\theta _{2})^{2}}{8T}-\frac{\theta _{1}^{2}\theta _{2}^{2}}{8T}+\frac{%
(\theta _{1}-\theta _{2})^{4}}{96T}\right)  \label{A3}
\end{equation}

We can neglect the term $(\theta _{1}-\theta _{2})^{4}/96T$ as will be seen
below.

As a result integral equation (\ref{I1}) reduces to
\begin{equation}
\int_{0}^{\pi }\sqrt{\frac{T\theta _{2}}{2\pi \theta _{1}}}\left( 1+\frac{T}{%
2\theta _{1}\theta _{2}}\right) e^{-\frac{(\theta _{1}-\theta _{2})^{2}}{8T}-%
\frac{\theta _{1}^{2}\theta _{2}^{2}}{8T}}\psi _{\alpha }(\theta
_{2})d\theta _{2}=\lambda _{\alpha }\psi _{\alpha }(\theta _{1})  \label{I2}
\end{equation}

The maximum of the expression in the exponent (saddle point) is at $\theta
_{2}=\theta _{1}$ (more exactly $\theta _{2}=\theta _{1}-\theta
_{1}^{3}+\ldots $, but it suffices to put $\theta _{2}=\theta _{1}$). Near
this saddle point we expand $\psi _{\alpha }(\theta _{2})$ as follows
\begin{equation}
\psi _{\alpha }(\theta _{2})=\psi _{\alpha }(\theta _{1})+(\theta
_{2}-\theta _{1})\psi _{\alpha }^{\prime }(\theta _{1})+\frac{(\theta
_{2}-\theta _{1})^{2}}{2}\psi _{\alpha }^{\prime \prime }(\theta
_{1})+\ldots   \label{psi1}
\end{equation}%
and
\begin{equation}
\sqrt{\frac{\theta _{2}}{\theta _{1}}}=\sqrt{1+\frac{\theta _{2}-\theta _{1}%
}{\theta _{1}}}=1+\frac{\theta _{2}-\theta _{1}}{2\theta _{1}}-\frac{(\theta
_{2}-\theta _{1})^{2}}{8\theta _{1}^{2}}+\ldots   \label{q}
\end{equation}

Let us introduce new scaled variables
\begin{eqnarray}
\theta _{2}-\theta _{1} &=&T^{1/2}x  \nonumber \\
\theta _{1} &=&T^{1/3}r  \label{xr}
\end{eqnarray}

Now $\psi _{\alpha }(\theta )\rightarrow \psi _{\alpha }(r)$ and
\begin{equation}
\psi _{\alpha }(\theta _{2})\rightarrow \psi _{\alpha }(r)+T^{1/6}x\psi
_{\alpha }^{\prime }(r)+\frac{T^{1/3}x^{2}}{2}\psi _{\alpha }^{\prime \prime
}(r)+\ldots  \label{psi2}
\end{equation}%
\begin{equation}
\sqrt{\frac{\theta _{2}}{\theta _{1}}}\rightarrow 1+\frac{T^{1/6}x}{2r}-%
\frac{T^{1/3}x^{2}}{8r^{2}}+O(T^{1/2})  \label{qq}
\end{equation}%
\begin{equation}
\exp \left( -\frac{(\theta _{1}-\theta _{2})^{2}}{8T}-\frac{\theta
_{1}^{2}\theta _{2}^{2}}{8T}\right) \rightarrow \exp \left( -\frac{x^{2}}{8}-%
\frac{T^{1/3}r^{4}}{8}\right)  \label{qqq}
\end{equation}

Summarizing all above we arrive at
\begin{equation}
\int_{-r/T^{1/6}}^{\pi /\sqrt{T}}\left( 1+\frac{T^{1/6}x}{2r}-\frac{%
T^{1/3}x^{2}}{8r^{2}}\right) \left( 1+\frac{T^{1/3}}{2r^{2}}\right) \left(
\psi _{\alpha }(r)+T^{1/6}x\psi _{\alpha }^{\prime }(r)+\frac{T^{1/3}x^{2}}{2%
}\psi _{\alpha }^{\prime \prime }(r)\right) e^{-x^{2}/8-T^{1/3}r^{4}/8}\frac{%
Tdx}{\sqrt{2\pi }}=\lambda _{\alpha }\psi _{\alpha }(r)  \label{I3}
\end{equation}

At $T\rightarrow 0$, we can change the limits in the integral to $[-\infty
,\infty ]$, then only even powers in $x$ gives contribution, so taking into
account only terms up to$~T^{1/3}$ we obtain
\begin{equation}
\int_{-\infty }^{\infty }\left[ \left( 1-\frac{T^{1/3}r^{4}}{8}+\frac{T^{1/3}%
}{2r^{2}}\right) \psi _{\alpha }+\frac{T^{1/3}x^{2}}{2}\left( \psi _{\alpha
}^{\prime \prime }+\frac{1}{r}\psi _{\alpha }^{\prime }(r)-\frac{1}{4r^{2}}%
\psi _{\alpha }\right) \right] e^{-x^{2}/8}\frac{Tdx}{\sqrt{2\pi }}=\lambda
_{\alpha }\psi _{\alpha }  \label{I4}
\end{equation}

After integration over $x$ we obtain a linear differential equation
\begin{equation}
2T\left( 1-\frac{T^{1/3}r^{4}}{8}+\frac{T^{1/3}}{2r^{2}}\right) \psi
_{\alpha }+4T^{4/3}\left( \psi _{\alpha }^{\prime \prime }+\frac{1}{r}\psi
_{\alpha }^{\prime }(r)-\frac{1}{4r^{2}}\psi _{\alpha }\right) =\lambda
_{\alpha }\psi _{\alpha }  \label{eq1}
\end{equation}%
and, finally,
\begin{equation}
-\psi _{\alpha }^{\prime \prime }-\frac{1}{r}\psi _{\alpha }^{\prime }+\frac{%
r^{4}}{16}\psi _{\alpha }=\varepsilon _{\alpha }\psi _{\alpha }  \label{eq2}
\end{equation}%
with
\begin{equation}
\varepsilon _{\alpha }=\frac{2T-\lambda _{\alpha }}{4T^{4/3}}  \label{e}
\end{equation}

Thus, we have got a Schr\"{o}dinger equation for a particle with $Z$
component of the angular momentum $l_{z}=0$ in 2D potential well $%
U(r)=r^{4}/16$. Normalization condition for $\psi _{\alpha }(r)$ is
\begin{equation}
\int_{0}^{\infty }\psi _{\alpha }(r)\psi _{\beta }(r)2\pi rdr=\delta
_{\alpha ,\beta }  \label{norma2}
\end{equation}

Numerical solution of Eq.(\ref{eq2}) gives the following lowest eigenvalues
(corresponding to the largest $\lambda $):
\begin{equation}
\varepsilon _{\alpha }=0.9305;3.78;7.44\ldots   \label{e_n}
\end{equation}

As was shown in Ref.\cite{Harada-Mikeska} the two-spin correlation function
can be expressed by the following integral
\begin{equation}
\left\langle \vec{S}_{1}\cdot \vec{S}_{1+n}\right\rangle =\frac{1}{\lambda
_{0}^{n-1}}\int_{0}^{\pi }d\theta _{n}\sin \theta
_{n}\prod_{l=1}^{n-1}d\theta _{l}\sin \theta _{l}\psi _{0}(\theta _{l})\psi
_{0}(\theta _{n})\left(
\begin{array}{cc}
0 & 1%
\end{array}%
\right) B(\theta _{1})H(\theta _{l},\theta _{l+1})B(\theta _{n})\left(
\begin{array}{c}
0 \\
1%
\end{array}%
\right)   \label{cor1}
\end{equation}%
where
\begin{equation}
B(\theta )=\left(
\begin{array}{cc}
\cos \theta /2 & \sin \theta /2 \\
-\sin \theta /2 & \cos \theta /2%
\end{array}%
\right)   \label{B}
\end{equation}%
\begin{equation}
H(\theta _{1},\theta _{2})=B(\theta _{1})\left(
\begin{array}{cc}
-\widetilde{A}(\theta _{1},\theta _{2}) & 0 \\
0 & A(\theta _{1},\theta _{2})%
\end{array}%
\right) B(\theta _{2})  \label{Hm}
\end{equation}%
and $\widetilde{A}(\theta _{1},\theta _{2})$ is given by Eq.(\ref{A1}) with $%
I_{0}(-z)$ replaced by $I_{1}(-z)$.

Using the asymptotic expansion of the Bessel function
\begin{equation}
I_{1}(-z)=-\frac{e^{z}}{\sqrt{2\pi z}}\left( 1-\frac{3}{8z}+O(z^{-2})\right)
\label{I1z}
\end{equation}%
we obtain
\begin{equation}
H(\theta _{1},\theta _{2})=A_{0}(\theta _{1},\theta _{2})\left(
\begin{array}{cc}
1-\frac{3T}{2\theta _{1}\theta _{2}} & \frac{\theta _{1}+\theta _{2}}{2} \\
-\frac{\theta _{1}+\theta _{2}}{2} & 1+\frac{T}{2\theta _{1}\theta _{2}}%
\end{array}%
\right)  \label{Hm2}
\end{equation}%
where
\begin{equation}
A_{0}(\theta _{1},\theta _{2})=\sqrt{\frac{T}{2\pi \theta _{1}\theta _{2}}}%
\exp \left( -\frac{(\theta _{1}-\theta _{2})^{2}}{8T}-\frac{\theta
_{1}^{2}\theta _{2}^{2}}{8T}\right)  \label{A0}
\end{equation}

The matrix $H(\theta _{1},\theta _{2})$ is not symmetric. Therefore, to
calculate $\left\langle \vec{S}_{1}\cdot \vec{S}_{1+n}\right\rangle $ it is
necessary to solve a pair of the integral equations%
\begin{eqnarray}
\int_{0}^{\pi }H(\theta _{1},\theta _{2})\vec{u}_{\alpha }(\theta _{2})\sin
\left( \theta _{2}\right) d\theta _{2} &=&\eta _{\alpha }\vec{u}_{\alpha
}(\theta _{1})  \label{in1} \\
\int_{0}^{\pi }H^{T}(\theta _{1},\theta _{2})\vec{v}_{\alpha }(\theta
_{2})\sin \left( \theta _{2}\right) d\theta _{2} &=&\eta _{\alpha }\vec{v}%
_{\alpha }(\theta _{1})  \label{in2}
\end{eqnarray}%
where $H^{T}(\theta _{1},\theta _{2})$ is transposed matrix $H(\theta
_{1},\theta _{2})$ and two-component vectors $\vec{u}_{\alpha }$ and $\vec{v}%
_{\alpha }$
\begin{equation}
\vec{u}_{\alpha }=\left(
\begin{array}{c}
u_{1,\alpha } \\
u_{2,\alpha }%
\end{array}%
\right) ,\qquad \vec{v}_{\alpha }=\left(
\begin{array}{c}
v_{1,\alpha } \\
v_{2,\alpha }%
\end{array}%
\right)  \label{uv}
\end{equation}%
satisfy orthonormality relations,
\begin{equation}
\int_{0}^{\pi }\vec{u}_{\alpha }^{T}(\theta )\vec{v}_{\beta }(\theta )\sin
\left( \theta \right) d\theta =\int_{0}^{\pi }\vec{v}_{\alpha }^{T}(\theta )%
\vec{u}_{\beta }(\theta )\sin \left( \theta \right) d\theta =\delta _{\alpha
,\beta }  \label{norma_uv}
\end{equation}

Then, the matrix $H(\theta _{1},\theta _{2})$ can be represented as
\begin{equation}
H(\theta _{1},\theta _{2})=\sum_{\alpha }\eta _{\alpha }\vec{u}_{\alpha
}(\theta _{1})\vec{v}_{\alpha }^{T}(\theta _{2})  \label{Hm3}
\end{equation}

At small$\ \theta _{1}$, $\theta _{2}$ Eqs.(\ref{in1}) and (\ref{in2})
reduce to
\begin{eqnarray}
\int\limits_{0}^{\pi }A_{0}(\theta _{1},\theta _{2})\left[ \left( 1-\frac{3T%
}{2\theta _{1}\theta _{2}}\right) u_{1,\alpha }(\theta _{2})+\theta
_{1}u_{2,\alpha }(\theta _{2})\right] \sin \left( \theta _{2}\right) d\theta
_{2} &=&\eta _{\alpha }u_{1,\alpha }(\theta _{1})  \label{in3} \\
\int\limits_{0}^{\pi }A_{0}(\theta _{1},\theta _{2})\left[ -\theta
_{1}u_{1,\alpha }(\theta _{2})+\left( 1+\frac{T}{2\theta _{1}\theta _{2}}%
\right) u_{2,\alpha }(\theta _{2})\right] \sin \left( \theta _{2}\right)
d\theta _{2} &=&\eta _{\alpha }u_{2,\alpha }(\theta _{1})  \label{in4}
\end{eqnarray}

Integrating these equations near the saddle point similar to Eqs.(\ref{I3}),(%
\ref{I4}), we get a pair of linear differential equation
\begin{eqnarray}
2T\left( 1-\frac{T^{1/3}r^{4}}{8}-\frac{3T^{1/3}}{2r^{2}}\right) u_{1,\alpha
}+4T^{4/3}\left( u_{1,\alpha }^{\prime \prime }+\frac{1}{r}u_{1,\alpha
}^{\prime }-\frac{1}{4r^{2}}u_{1,\alpha }\right) +2T^{4/3}ru_{2,\alpha }
&=&\eta _{\alpha }u_{1,\alpha }  \label{eq_u1} \\
2T\left( 1-\frac{T^{1/3}r^{4}}{8}+\frac{T^{1/3}}{2r^{2}}\right) u_{2,\alpha
}+4T^{4/3}\left( u_{2,\alpha }^{\prime \prime }+\frac{1}{r}u_{2,\alpha
}^{\prime }-\frac{1}{4r^{2}}u_{2,\alpha }\right) -2T^{4/3}ru_{1,\alpha }
&=&\eta _{\alpha }u_{2,\alpha }  \label{eq_u2}
\end{eqnarray}%
and, finally,
\begin{eqnarray}
-u_{1,\alpha }^{\prime \prime }-\frac{1}{r}u_{1,\alpha }^{\prime }+\frac{1}{%
r^{2}}u_{1,\alpha }+\frac{r^{4}}{16}u_{1,\alpha }+\frac{r}{2}u_{2,\alpha }
&=&\mu _{\alpha }u_{1,\alpha }  \label{eq_u11} \\
-u_{2,\alpha }^{\prime \prime }-\frac{1}{r}u_{2,\alpha }^{\prime }+\frac{%
r^{4}}{16}u_{2,\alpha }-\frac{r}{2}u_{1,\alpha } &=&\mu _{\alpha
}u_{2,\alpha }  \label{eq_u12}
\end{eqnarray}%
where
\begin{equation}
\mu _{\alpha }=\frac{2T-\eta _{\alpha }}{4T^{4/3}}  \label{mu}
\end{equation}

A few lowest eigenvalues of Eqs.(\ref{eq_u11}) and (\ref{eq_u12}) are
\begin{equation}
\mu _{\alpha }=1.4113;1.83;3.98\ldots   \label{mu_n}
\end{equation}

For $\vec{v}_{\alpha }$ similar procedure gives
\begin{eqnarray}
-v_{1,\alpha }^{\prime \prime }-\frac{1}{r}v_{1,\alpha }^{\prime }+\frac{1}{%
r^{2}}v_{1,\alpha }+\frac{r^{4}}{16}v_{1,\alpha }-\frac{r}{2}v_{2,\alpha }
&=&\mu _{\alpha }v_{1,\alpha }  \label{eq_v1} \\
-v_{2,\alpha }^{\prime \prime }-\frac{1}{r}v_{2,\alpha }^{\prime }+\frac{%
r^{4}}{16}v_{2,\alpha }+\frac{r}{2}v_{1,\alpha } &=&\mu _{\alpha
}v_{2,\alpha }  \label{eq_v2}
\end{eqnarray}

It follows from Eqs.(\ref{eq_u11})-(\ref{eq_u12}) and (\ref{eq_v1})-(\ref%
{eq_v2}) that the functions $\vec{v}_{\alpha }$ is connected with $\vec{u}%
_{\alpha }$ by the relations $v_{1,\alpha }=-u_{1,\alpha }$, $v_{2,\alpha
}=u_{2,\alpha }$ and, therefore, normalization condition (\ref{norma_uv})
transforms to
\begin{equation}
T^{2/3}\int_{0}^{\infty }\left( u_{2,\alpha }u_{2,\beta }-u_{1,\alpha
}u_{1,\beta }\right) rdr=\delta _{\alpha ,\beta }  \label{norma_u}
\end{equation}

Using Eqs.(\ref{Hm3}) and (\ref{norma_uv}) we obtain the correlation
function (\ref{cor1}) in a form
\begin{equation}
\left\langle \vec{S}_{1}\cdot \vec{S}_{1+n}\right\rangle =\sum_{\alpha
}y_{\alpha }^{n-1}f_{\alpha }^{2}  \label{cor2}
\end{equation}%
where $y_{\alpha }=\eta _{\alpha }/\lambda _{0}$ and
\begin{equation}
f_{\alpha }=\int_{0}^{\pi }\psi _{0}(\theta )u_{2,\alpha }(\theta )\sin
\left( \theta \right) d\theta =T^{2/3}\int_{0}^{\infty }\psi
_{0}(r)u_{2,\alpha }(r)rdr  \label{f}
\end{equation}

At $T\rightarrow 0$
\begin{equation}
y_{\alpha }=\frac{2T-4T^{4/3}\mu _{\alpha }}{2T-4T^{4/3}\varepsilon _{0}}%
\approx 1-2T^{1/3}\left( \mu _{\alpha }-\varepsilon _{0}\right)   \label{y}
\end{equation}%
and\ the correlation function becomes%
\begin{equation}
\left\langle \vec{S}_{1}\cdot \vec{S}_{1+n}\right\rangle =\sum_{\alpha
}f_{\alpha }^{2}\exp [-2T^{1/3}\left( \mu _{\alpha }-\varepsilon _{0}\right)
(n-1)]  \label{cor3}
\end{equation}

According to Eq.(\ref{cor3}) the correlation length $\xi $ at $T\rightarrow 0
$ is
\begin{equation}
\xi =\frac{1}{2\left( \mu _{0}-\varepsilon _{0}\right) T^{1/3}}=\frac{1.04}{%
T^{1/3}}  \label{xi}
\end{equation}

Now we are ready to calculate the magnetic susceptibility at $T\rightarrow 0$%
, which is%
\begin{equation}
\chi =\frac{1}{3TN}\sum_{n}\left\langle \vec{S}_{1}\cdot \vec{S}%
_{1+n}\right\rangle =\frac{1}{3T}(1+2\sum_{\alpha }\frac{f_{\alpha }^{2}}{%
1-y_{\alpha }})=\frac{1}{3T}+\frac{1}{3T^{4/3}}\sum_{\alpha }\frac{f_{\alpha
}^{2}}{\mu _{\alpha }-\varepsilon _{0}}  \label{chi}
\end{equation}

Now we see that $f_{\alpha }^{2}$ and $(\mu _{\alpha }-\varepsilon _{0})$
depends on the solutions of differential equations which are independent of $%
T$. So, the sum in $\chi $ gives numerical constant
\begin{equation}
\sum_{\alpha }\frac{f_{\alpha }^{2}}{\mu _{\alpha }-\varepsilon _{0}}=3C
\label{C}
\end{equation}

Therefore, the low-temperature susceptibility behaves as
\begin{equation}
\chi =\frac{C\left\vert J_{1}^{1/3}\right\vert }{T^{4/3}}  \label{chi_f}
\end{equation}

Numerical calculations gives for the constant $C$ the value $C\approx 1.07$.
Thus, the critical exponent for the susceptibility at the transition point
is $4/3$ and that for the correlation length is $1/3$.

\end{document}